\documentclass{iaus}
\usepackage[dvips]{graphicx}

  \checkfont{eurm10}
  \iffontfound
    \IfFileExists{upmath.sty}
      {\typeout{^^JFound AMS Euler Roman fonts on the system,
                   using the 'upmath' package.^^J}%
       \usepackage{upmath}}
      {\typeout{^^JFound AMS Euler Roman fonts on the system, but you
                   dont seem to have the}%
       \typeout{'upmath' package installed. iaus.cls can take advantage
                 of these fonts,^^Jif you use 'upmath' package.^^J}%
      }
  \else
  \fi

  \checkfont{msam10}
  \iffontfound
    \IfFileExists{amssymb.sty}
      {\typeout{^^JFound AMS Symbol fonts on the system, using the
                'amssymb' package.^^J}%
       \usepackage{amssymb}%

      }{}
  \fi

  \IfFileExists{amsbsy.sty}
    {\typeout{^^JFound the 'amsbsy' package on the system, using it.^^J}%
     \usepackage{amsbsy}}
    {\providecommand\boldsymbol[1]{\mbox{\boldmath $##1$}}}

\newsavebox{\astrutbox}
\sbox{\astrutbox}{\rule[-5pt]{0pt}{20pt}}

\title[Outskirts of Galaxy Clusters: intense life in the suburbs]
      {Radio halos in merging clusters of galaxies}

\author[S.Giacintucci {\it et al.\/}]%
{S.Giacintucci$^{1,2}$, T.Venturi$^3$, S.Bardelli$^2$, G.Brunetti$^3$,
 D.Dallacasa$^{1,3}$, P.Rao$^4$, E.Zucca$^2$}

\affiliation{$^1$Dipartimento di Astronomia, Univ. Bologna, via Ranzani 1, I- 40127 Bologna, Italy \\[\affilskip]
$^2$ Osservatorio astronomico, via Ranzani 1, I- 40127 Bologna, Italy \\[\affilskip]
$^3$ Istituto di Radioastronomia - CNR, via Gobetti 101, I-40129 Bologna, Italy \\[\affilskip]
$^4$ National Centre for Radio Astrophysics, Pune University Campus, India }

\pubyear{2004}
\volume{195}
\pagerange{1--8}
\date{?? and in revised form ??}
\jname{Outskirts of Galaxy Clusters: intense life in the suburbs}
\editors{A. Diaferio, ed.}
\begin{document}

\maketitle

\begin{abstract}
We present the preliminary results of 235 MHz, 327 MHz and 610 MHz observations
 of the galaxy cluster A3562 in the core of the Shapley Concentration. The purpose
 of these observations, carried out with the Giant Metrewave Radio Telescope
(GMRT, Pune, India), was to study the radio halo located at the centre of A3562
 and determine the shape of its radio spectrum at low frequencies, in order to 
understand the origin of this source. In the framework of the re--acceleration 
model, the preliminary analysis of the halo spectrum suggests that we are 
observing a young source (few $10^8$ yrs) at the beginning of the 
re--acceleration phase.
\end{abstract}

\firstsection 
\section{Introduction}
A number of X--ray luminous galaxy clusters show large--scale synchrotron 
radio emission associated to the intracluster medium. These diffuse 
sources are know as radio halos, when they are located at the centre of the 
hosting cluster and show low or negligible polarization, and radio relics, 
when found in the cluster outskirts and highly polarized.

\noindent Both halos and relics have low surface brightness, large linear size
 (from $\sim$0.5 Mpc to more than 1 Mpc) and steep integrated spectra, i.e. 
$\alpha >1$ (S $\propto$ $\nu^{-\alpha}$).

\noindent Radio halos and relics represent the most striking evidence for $\mu$G 
magnetic fields on cluster scale and relativistic electrons diffused within the 
whole cluster volume (for a recent review see Giovannini \& Feretti 2002).

\noindent The existence of this class of radio sources is believed to be connected 
to cluster mergers, since thus far they have only been found in clusters with 
significant signs of a current or recent merging event. 
In particular the leading hypothesis for the origin of the observed radio 
emission from these objects is a re--acceleration process, probably via 
turbolence powered by cluster mergers, of a population of relatively
low energy ($\gamma \sim 10^3$) electrons, initially injected in the
intracluster medium (two-phase model, Brunetti et al. 2001).

\section{The radio halo in A3562}
The cluster A3562 is the easternmost component of the A3558 complex, a chain
of three clusters and two SC groups which form a single physically connected
structure, whose complex dynamical stage of merging is supported by a large
amount of observational data in all bands (see for instance Ettori et al. 1997,
Ettori et al. 2000). 

\noindent The existence of a radio halo at the A3562 centre was confirmed by 
deep radio observations at 1.4 GHz carried out with VLA (Venturi et al. 2003).

\noindent In the framework of the re--acceleration model, the halo in A3562 seems to 
play a very special role, since its largest linear size (LLS $\sim$ 600 kpc)
 and 1.4 GHz radio power (P$_{1.4GHz}$ = 2.14$\times 10^{23}$ W Hz$^{-1}$) 
are among the lowest values found for this class of radio sources.

\noindent The halo nicely fits the correlations existing between the halo radio power 
and the cluster X--ray luminosity and temperature (Bacchi et al. 2003), 
extending them to lower values for all quantities involved.

\noindent According to the two-phase model, the halo in A3562 could be interpreted 
either as the result of a low efficiency re-acceleration process or as a 
young source at the beginning of the re-acceleration phase. It has been 
established that the
head--tail radio galaxy J1333--3141, located within the halo emission, has
deposited in the intracluster medium a number of electrons high enough to 
feed the halo if it has been active for a considerable fraction of its 
crossing time (t$_{cross}\sim 6 \times 10^8$ yrs). This result seems to
 suggest that the halo at the centre of A3562 is young as compared to the
 other radio halos 
known to date (Venturi et al. 2003). However only low frequencies observations
allow to discriminate between the two possibilities.

\section{Low frequencies observations of the halo in A3562}
With the purpose of studing the morphology of the halo in
A3562 and determine the shape of its synchrotron spectrum at
low frequencies, we carried out observations at 235 MHz, 327 MHz and
610 MHz with the Giant Metrewave Radio Telescope (GMRT, Pune, India).
Thanks to the u--v coverage of this instrument, a range of resolutions
at all frequencies (from few arcsec to tens of arcsec) is allowed. This
ensures an accurate determination of the total radio flux density of the
halo, after a careful subtraction of the individual point sources embedded
in the halo emission.

\noindent In Figure 1 we report the image of the radio halo in A3562 at 327 MHz overlaid 
on the optical DSS--2 image, and in Figure 2 we present the 610 MHz image of the
halo overlaid on the X--ray XMM image.

\begin{figure}
\begin{center}
  \caption{327 MHz GMRT image of the radio halo in A3562 and the south--western
extended emission in the direction of the SC 1329--313 group, overlaid on
the optical DSS--2 image. The restoring beam
is $41.98^{\prime \prime} \times 35.13^{\prime \prime}$. The rms in the image
is $\sim 0.25$ mJy/beam and the lowest contour is 0.8 mJy/beam.}
\end{center}
\end{figure}

\begin{figure}
\caption{610 MHz GMRT image of the radio halo in A3562 overlaid on the X--ray
XMM image. 
The restoring beam
is $41.98^{\prime \prime} \times 35.13^{\prime \prime}$. The rms in the image
is $\sim 0.15$ mJy/beam and the lowest contour is 0.45 mJy/beam.}
\end{figure}

\noindent An extended radio emission, located south--west of the halo, 
surrounds the radio source J1332--3146 (Figure 1), identified with 
the brightest galaxy in the SC 1329--313 group (Giacintucci et al. 2004).
The nuclear component, detected at 1.4 GHz, disappears at low frequency. 
This seems to rule out any connection between the low brightness extended
emission and an active nucleus, suggesting that its origin is most likely
due to cluster merger re--acceleration.

\noindent The preliminary results of our low frequency data suggest that
the spectrum of the radio halo is steep up to $\sim$ 610 MHz, with 
$\alpha_{610 MHz}^{1.4 GHz} \sim 1.9$, and it flattens beyond 610 MHz,
with $\alpha_{235 MHz}^{610 MHz} \sim 1$. This seems to be in good 
agreement with a very recent re--acceleration phase, supporting 
the idea that the radio halo in A3562 is young and that its existence 
is strongly connected to the ongoing major merger in the A3558 complex
(Venturi et al. in prep).

\end{document}